\documentclass[10pt,journal,cspaper,compsoc]{IEEEtran}
\ifCLASSOPTIONcompsoc
  \usepackage[nocompress]{cite}
\else
  \usepackage{cite}
\fi
\ifCLASSINFOpdf
  \usepackage{graphicx}
\else
\fi
\usepackage{epstopdf} 
\usepackage[cmex10]{amsmath}
\usepackage[linesnumbered]{algorithm2e}
\usepackage{color}
\ifCLASSOPTIONcompsoc
\usepackage[tight,normalsize,sf,SF]{subfigure}
\else
\usepackage[tight,footnotesize]{subfigure}
\fi
\usepackage{caption}
\begin{document}
%
% paper title
% can use linebreaks \\ within to get better formatting as desired
\title{Crowd Behavior Simulation with Emotional Contagion in Unexpected Multi-hazard Situations}
%
%
% author names and IEEE memberships
% note positions of commas and nonbreaking spaces ( ~ ) LaTeX will not break
% a structure at a ~ so this keeps an author's name from being broken across
% two lines.
% use \thanks{} to gain access to the first footnote area
% a separate \thanks must be used for each paragraph as LaTeX2e's \thanks
% was not built to handle multiple paragraphs
%
%
%\IEEEcompsocitemizethanks is a special \thanks that produces the bulleted
% lists the Computer Society journals use for "first footnote" author
% affiliations. Use \IEEEcompsocthanksitem which works much like \item
% for each affiliation group. When not in compsoc mode,
% \IEEEcompsocitemizethanks becomes like \thanks and
% \IEEEcompsocthanksitem becomes a line break with idention. This
% facilitates dual compilation, although admittedly the differences in the
% desired content of \author between the different types of papers makes a
% one-size-fits-all approach a daunting prospect. For instance, compsoc
% journal papers have the author affiliations above the "Manuscript
% received ..."  text while in non-compsoc journals this is reversed. Sigh.

 \author{Mingliang~Xu,
         Xiaozheng~Xie,
         Pei~Lv,
		 Jianwei Niu,
         Hua~Wang \\
	 Chaochao~Li,
     Ruijie Zhu,
         Zhigang~Deng
         and Bing~Zhou %<-this % stops a space
  \IEEEcompsocitemizethanks{
  \IEEEcompsocthanksitem Mingliang Xu, Pei Lv, Chaochao Li, Ruijie Zhu and Bing Zhou are with Center for Interdisciplinary Information Science Research, ZhengZhou University, 450000. Dr.Pei Lv is the corresponding author.
  \IEEEcompsocthanksitem Xiaozheng Xie, Jianwei Niu is with State Key Laboratory of Virtual Reality Technology and Systems, Beihang University,100091
  \IEEEcompsocthanksitem Hua Wang is with School of Computer and Communication Engineering of Zhengzhou University of Light Industry, 450000.
  \IEEEcompsocthanksitem Zhigang Deng is with Department of Computer Science, University of Houston, Houston, TX, 77204-3010.\protect\\
  E-mail: \{iexumingliang, ielvpei, zhuruijie, iebzhou\} @zzu.edu.cn; \hfil\break xiexzheng@163.com;niujianwei@buaa.edu.cn; wanghua@zzuli.edu.cn; zzulcc@gs.zzu.edu.cn; \hfil\break zdeng4@uh.edu
 }% <-this % stops a space
 \thanks{}
 }

\IEEEcompsoctitleabstractindextext{%
\begin{abstract}
Numerous research efforts have been conducted to simulate crowd movements, while relatively few of them are specifically focused on multi-hazard
situations. In this paper, we propose a novel crowd simulation method by modeling the generation and contagion of panic emotion under multi-hazard
circumstances. In order to depict the effect from hazards and other agents to crowd movement, we first classify hazards into different types
(transient and persistent, concurrent and non-concurrent, static and dynamic) based on their inherent characteristics. Second, we introduce the
concept of perilous field for each hazard and further transform the critical level of the field to its invoked-panic emotion. After that, we propose
an emotional contagion model to simulate the evolving process of panic emotion caused by multiple hazards. Finally, we introduce an Emotional
Reciprocal Velocity Obstacles (ERVO) model to simulate the crowd behaviors by augmenting the traditional RVO model with emotional contagion,
which for the first time combines the emotional impact and local avoidance together. Our experiment results demonstrate that the overall approach
is robust, can better generate realistic crowds and the panic emotion dynamics in a crowd. Furthermore, it is recommended that our method can be
applied to various complex multi-hazard environments.
\end{abstract}

% IEEEtran.cls defaults to using nonbold math in the Abstract.
% This preserves the distinction between vectors and scalars. However,
% if the journal you are submitting to favors bold math in the abstract,
% then you can use LaTeX's standard command \boldmath at the very start
% of the abstract to achieve this. Many IEEE journals frown on math
% in the abstract anyway. In particular, the Computer Society does
% not want either math or citations to appear in the abstract.

% Note that keywords are not normally used for peer review papers.
}

% make the title area
\maketitle

% To allow for easy dual compilation without having to reenter the
% abstract/keywords data, the \IEEEcompsoctitleabstractindextext text will
% not be used in maketitle, but will appear (i.e., to be "transported")
% here as \IEEEdisplaynotcompsoctitleabstractindextext when compsoc mode
% is not selected <OR> if conference mode is selected - because compsoc
% conference papers position the abstract like regular (non-compsoc)
% papers do!
\IEEEdisplaynotcompsoctitleabstractindextext
% \IEEEdisplaynotcompsoctitleabstractindextext has no effect when using
% compsoc under a non-conference mode.

% For peer review papers, you can put extra information on the cover
% page as needed:
% \ifCLASSOPTIONpeerreview
% \begin{center} \bfseries EDICS Category: 3-BBND \end{center}
% \fi
%
% For peerreview papers, this IEEEtran command inserts a page break and
% creates the second title. It will be ignored for other modes.
\IEEEpeerreviewmaketitle

% Computer Society journal papers do something a tad strange with the very
% first section heading (almost always called "Introduction"). They place it
% ABOVE the main text! IEEEtran.cls currently does not do this for you.
% However, You can achieve this effect by making LaTeX jump through some
% hoops via something like:
%
%\ifCLASSOPTIONcompsoc
%  \noindent\raisebox{2\baselineskip}[0pt][0pt]%
%  {\parbox{\columnwidth}{\section{Introduction}\label{sec:introduction}%
%  \global\everypar=\everypar}}%
%  \vspace{-1\baselineskip}\vspace{-\parskip}\par
%\else
%  \section{Introduction}\label{sec:introduction}\par
%\fi
%
% Admittedly, this is a hack and may well be fragile, but seems to do the
% trick for me. Note the need to keep any \label that may be used right
% after \section in the above as the hack puts \section within a raised box.

% The very first letter is a 2 line initial drop letter followed
% by the rest of the first word in caps (small caps for compsoc).
%
% form to use if the first word consists of a single letter:
% \IEEEPARstart{A}{demo} file is ....
%
% form to use if you need the single drop letter followed by
% normal text (unknown if ever used by IEEE):
% \IEEEPARstart{A}{}demo file is ....
%
% Some journals put the first two words in caps:
% \IEEEPARstart{T}{his demo} file is ....
%
% Here we have the typical use of a "T" for an initial drop letter
% and "HIS" in caps to complete the first word.

\section{Introduction} 

\IEEEPARstart{T}he advances in the study of typical crowd behaviors (such as stampede 
incidents and terrorist attacks) in various domains including psychology, 
security management, and computer science, have pointed out that simulating 
both the sentimental state evolution and decision-making of a crowd under different 
circumstances is an efficient way to show inherent laws of nature \cite{1}. This problem has been considered as a system that as a class of multi-input multi-output systems in the non-strict feedback structure \cite{45}. As 
a result, it is important to accurately model both the simulation environment and 
emotional contagion among individuals for realistic crowd simulation. 

Recent research efforts of crowd simulation in emergency circumstances have been mostly 
focused on those situations where there is only one hazard in the area of interest \cite{2,3,5,6,7,8,9}. However, in some real-world cases, multiple hazards may 
occur in the same area over a period of time, such as the two sequential bombing attacks in 
Boston in 2013. Traditional crowd simulation algorithms 
with a single hazard in the scenario cannot be applied to these 
cases directly because of the following reasons: 

1) A multi-hazard scenario, including different types of hazards, different critical 
levels of hazards, dynamic changes of hazards, various evacuation strategies, and so on, is more complex than the case with a single hazard. 
The traditional single-hazard models are very difficult to handle all the above factors in a unified way.

2) The emotional contagion in multi-hazard environment is a complex 
combining process of emotional spreading, concerning both direct effects from 
hazards and indirect effects from neighboring individuals. However, existing emotional contagion models are mainly designed for single-hazard scenes and cannot 
be applied to multi-hazard scenes directly.

3) Traditional multi-agent navigation algorithms, like Reciprocal Velocity Obstacles (RVO)~\cite{8}, have not considered the emotion of individuals, which means they are short of the 
mechanism to deal with the conflict between obstacle avoidance and panic escaping. 
Therefore, the simulation results under multi-hazard circumstance by these 
algorithms appear less realistic.

In order to tackle the above challenges, in this paper, we propose a novel 
multi-hazard scene model to describe different effects of various types of 
hazards, which is mainly applied to fire and explosion 
situations. In this model, the hazards are classified into six different 
types according to three kinds of inherent attributes: durations, time of 
occurence and dynamics. Based on the definitions of these hazards,  we 
further propose the concept of perilous field and a conversion function to 
map the criticality of the perilous field to the emotion of individuals. It 
is noteworthy that emotion in this paper mainly refers to the panic mood of 
individuals in emergency situations. 
 
In order to depict the complex process of panic spreading, we put forward a new emotional contagion model 
specially designed for multi-hazard situations by combining panic emotions from different hazards and individuals. 
Finally, an Emotional Reciprocal Velocity Obstacles(ERVO) model, inspired from the traditional RVO model, 
is proposed to drive the crowd movement. Different from the existing RVO model, the ERVO model integrates the 
emotional effect into velocity decision for the first time.  

The contributions of this paper are:

\begin{itemize}
 \item We propose a novel multi-hazard scene model for the description of emergency fire and explosion situations, 
 containing six different types of hazards with their dynamic changing process and an unified criticality conversion function.
 
 \item We propose a new emotional contagion model in multi-hazard scenarios, which 
 combines different emotional effects from hazards and individuals in a crowd.
  
 \item We propose a novel crowd behavior simulation method, the ERVO to simulate how people under a panic 
 mode choose their paths to safe places or planned goals in a realistic way. 
 
\end{itemize}

The rest of this paper are organized as follows. Background and related work are reviewed in Section 2. The overview 
of our work is introduced in Section 3. The definition of multiple types of hazards and emergency scenes are described in Section 4. 
The emotional contagion process is explained in detail in Section 5. The simulation method of crowd movement 
is described in Section 6. Our experiments are presented in Section 7. Finally, this paper is concluded in Section 8.

\section{Related Work}

Although numerous research efforts have been conducted to simulate crowd 
movements, relatively little literature has been specifically focused on emergency 
evacuation simulation involved with multiple hazards. In this section, we 
will mainly review recent works that are clearly related to our work. For 
more comprehensive review on crowd simulation techniques, please refer to 
\cite{35}. 

\subsection{Crowd evacuation with social or physical model}
One kind of important crowd movement scenarios is to simulate the emergency 
evacuation. Helbing et al.\cite{2} employ the social force model, combined 
with social psychology and physics models for the first time, to describe 
the panic behavior in evacuation. After that, the lattice gas 
model~\cite{13}, multi-grid model \cite{14}, agent-based model \cite{5}, 
virtual hindrance model \cite{15}, etc., have also been proposed to describe 
the dynamical behaviors of the emergency crowd. The commonness among these methods 
is that they choose some typical characteristics of the crowd first, 
and then use corresponding models to describe different evacuation behaviors. 
Other studies considering more factors in crowd evacuation process, Narain et al. \cite{3} simulate the clustering behaviors of a high 
density crowd in a combined macro-micro perspective. Funge et al. \cite{16} 
put forward a cognitive model to direct autonomous characters to perform specific tasks, 
which outperforms many traditional behaviors models. Durupinar et al. \cite{17} analyze the impact of psychological 
factors on the crowd movement from the perspective of social psychology. 
Lai et al. \cite{46} aim at a problem of adaptive quantized control for a class of uncertain nonlinear systems preceded by asymmetric actuator backlash, 
which is similar with our motion analysis with agents in unexpected situations.
Wang et al. \cite{Trending7797248} propose a semantic-level crowd evaluation metric, which 
analyze the semantic information between real and simulated data.
Basak et al.\cite{Using1802} validate and optimize crowd simulation by using a data-driven approach, which proves the parameters 
learned from the real videos can better represent the common traits of incidents when simulation. 
Oguz et al. \cite{1} use continuous dynamic model, to simulate the movements 
of agents in outdoor emergency situations successfully. In this paper, our 
crowd behavior model mainly focuses on the micro-level behavior simulation. According to 
different multi-hazard environments, we divide the crowd movement into 
various cases and design crowd behaviors for each case specially.

\subsection{Crowd simulation with psychological model}
In the real world, emotional state of an individual plays a vital role 
in his/her decision-making, which fundamentally determines his/her movements 
at each time step~\cite{19,20}. Therefore, many recent works start to 
consider the psychological factors of agents, especially during the simulating process 
of crowd movement ~\cite{21}. Belkaid et al. \cite{Autonomous2018} stress the important 
role that emotional modulation plays on behavior organization by analyzing the relationships between emotion and cognition. Bosse et al.\cite{18} propose the absorption 
model based on the heat dissipation theory in thermodynamic, which embodies 
the role of authority figures in the process of emotional contagion. 
Tasi et al. \cite{36} devise a multi-agent evacuation simulation tool ESCAPES, where an 
agent will accept the emotion of other agents who has the strongest mood or has 
special identity. Le et al. \cite{23} propose an agent-based evacuation model by 
considering emotion propagation among individuals to make the simulation 
more realistic. Lhommet et al. \cite{37} also propose a computational model of emotional 
contagion based on individual personality and relationships. Durupinar et al. \cite{17} 
create a system that enables the specification of different crowd types 
ranging from audiences to mobs based on a computational mapping from the OCEAN 
personality traits to emotional contagion. Tsai et al. \cite{39} combine the 
dynamics-based and epidemiologicial-based models to describe the dynamics of 
emotional spreading from the perspective of social psychology. Fu et al. \cite{11} use  
a modified SIR model, originally proposes in \cite{40}, to model the emotion 
evolving in the process of emergency crowd movement. The work in \cite{38} 
proposes a stress model to realize the interactive simulation of dynamic 
crowd behaviors. Although stress is similar to our panic emotion in terms of 
the impact on crowd behaviors, there are still some inherent differences. 
For one certain crowd scene, they mainly model one type of stress in 
it and the stress of external environment on individuals. The mutual influence impact among different individuals is ignored. In addition, their model only 
focuses on the changes of individuals' velocities caused by the magnitude of stress. By contrast, in our paper, the emotional state of agents in emergency 
situations is mainly the panic emotion. Due to different emotional spreading and reception for various agents, we analyze the emotional contagion by involving the personality factors. Since the panic effect is not only 
coming from various hazards but also from neighboring individuals, a new 
micro-continuous emotion contagion model is designed.

\subsection{Crowd path planning}
Generally, path planning can be regarded as the multi-objective optimization \cite{wang2018decomposition} and local information interaction \cite{WangModel, zhang2018consensus} problems. 
In the process of crowd evacuation, an individual's action decision \cite{Park2004Toward, Cui2013Tracking, Garc2017Motion,zelek2000local-global,qu2009simulating,park2008memory-based,Event8360958} is 
dependent on the evacuating directions of nearby agents, the locations of 
hazards, and the obstacles in the scene.

Some researches develop a variety of methods to avoid the collision problem through the calculation of possible 
positions of individuals at the next time step \cite{41,26}. On the premise of 
collision avoidance, Kluge and Prassler \cite{27} use a local obstacle avoidance approach, combined with individual's emotion states to calculate the movements of 
agents iteratively. Van den Berg et al. \cite{8} propose the well-known Reciprocal Velocity 
Obstacles (RVO) model to drive the multi-agent navigation without collision. Concretely, the reactive behavior of one agent at each time step depends on the behaviors of all the other agents. In their method, a collision-avoidance velocity for each agent is chosen by taking into account the positions and reciprocal velocities of all agents in the scenario.
By constructing visual trees, Belkhouche \cite{28} proposes a shortest path without conflict. Guy et al. \cite{9} propose an optimization 
method for collision avoidance on the basis of the RVO model for real-time 
simulation of large-scale crowd movement. In addition, they also propose an energy-saving 
simulation method with the minimum energy consumption as the guidelines \cite{29}.  
Furthermore, a series of path planning and 
navigation algorithms \cite{31,32,33,34} are also described in mass population under complex background. In 
this paper, we enhance the traditional RVO model with emotional contagion in 
multi-hazard circumstances. Panic is used to describe the emotional 
state of each agent, which is changed dynamically and affect the 
behaviors of individuals.

\section{System Overview}

\begin{figure*}[t] \begin{centering} 
  \centering 
  \includegraphics[width=0.9\linewidth]{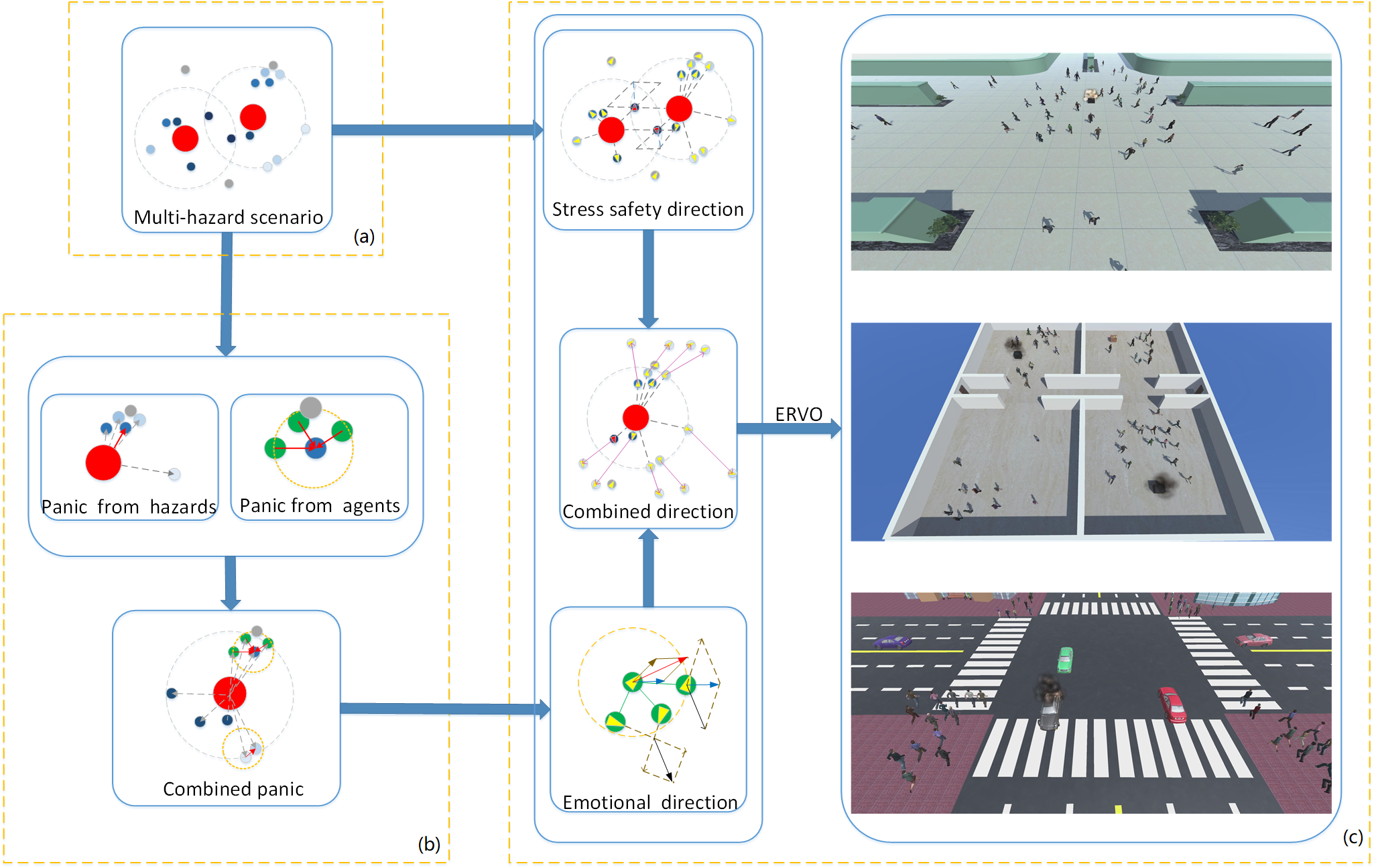}
  \centering
  \caption{The framework of crowd behavior simulation in multi-hazard situations, consisting of three parts: (a) the 
estimation of crowd panic in multi-hazard environment, (b) panic propagation in emergency situation and (c) 
the impact on crowd movement from panic emotion. The red solid circle represents the hazard in our 
circumstance, different blue solid circles in (a) represent agents with different panic emotion values. 
The darker the color, the greater the panic emotion. In (b), the panic emotion of one agent (blue solid circle) 
is affected by other agents (green solid circles) in its perceiving range. Stress safety directions, emotional 
directions and combined directions of agents are annotated by yellow arrows shown in (c).  
}
  \label{fig:1}
  \end{centering} 
\end{figure*}

As shown in Figure \ref{fig:1}, the main methodology of this work is divided 
into three parts: multi-hazard environment modeling in Section 
\ref{multihazard}; emotional contagion process under multi-hazard situations 
in Section \ref{emotional-contagion}; crowd behavior simulation based on 
emotional reciprocal velocity obstacle in Section \ref{evacuating}. 

Specifically, in order to simulate the crowd behavior in multi-hazard 
situations realistically, we analyze different types of 
hazards according to their properties, the time of occurrence and duration. 
After that, we propose a perilous field consisting of multiple hazards and 
define a conversion function to map the intensity of danger to panic emotion. Besides the 
direct effects from hazards, panic propagations also exist among different 
agents in emergency scenes. So we build an emotional contagion model 
(ECM) to handle the above cases. The ECM computes the panic emotion of each 
agent in the dangerous field according to the distance between this agent 
and the hazards using the above conversion function. At the same time, the ECM 
accumulates the contagious panic emotion from other agents to obtain the 
final emotion of each agent. To realize multi-agent navigation with panic 
emotion under multi-hazard situations, we propose an ERVO model to simulate 
the crowd behaviors. The major contribution of ERVO is a new mechanism of 
velocity decision by integrating both the traditional RVO and panic emotion.

\section{Multi-Hazard Environment Modeling}
\label{multihazard}

The characteristics of complexity, interactivity and time-varying make crowd 
behavior simulation challenging, especially in multi-hazard environments. In 
order to achieve realistic simulation results, we first need to model 
multi-hazard simulation environment quantitatively.

According to their durations, we divide hazards into two different 
types: transient and persistent. The former only lasts for a moment, while 
the latter lasts for relatively long time. Both of them would cause drastic 
changes to the psychological state of a crowd, and individuals in the 
dangerous area would respond immediately. The difference between them is that a transient 
hazard only threats those individuals at the time when it is happening. Once 
it disappears, the threat will also disappear immediately. By contrast, a 
persistent hazard will continue to impact those individuals in the dangerous 
area during its existence.

According to their generation time, we divide hazards into concurrent and 
non-concurrent. Specifically, when some hazards occur concurrently, their 
influences on neighboring agents can be treated as a single one. These influences should be accumulated together. 
For non-concurrent hazards, we need to consider the status of the crowd each time when a new hazard happens. 
If an agent has already been affected by other hazards before or has its own emotion, the new effect needs to be accumulated. 

More importantly, the static and dynamic characteristics of hazards also play vital effects on 
the crowd movement in complex situations. Based on this fact, we classify the hazards with fixed 
position and influence radius as static ones. Other cases, such as fixed position with variable 
influence radius, variable position with fixed or variable influence radius are regarded as dynamic 
hazards. For dynamic hazards, they may have different states over the time, which determine their position and
area of influence dynamically.

The above six basic types of hazards have obviously different impacts on the 
crowd movement. Realistic multi-hazard scenarios usually consist of these 
basic types and their combinations. 

After analyzing these hazards qualitatively, we give quantitative descriptions for them. 
We first define a perilous field as the circular area with the hazard position as the center 
and a radius. Each agent is aware of the existence of hazards in the 
scene through self-perception or neighbor contagion. The influence of danger
is limited in space: the farther the distance to the hazard, the weaker 
influence to the crowd. 
For different types of hazards, due to the uncertainty of their location and 
range, new perilous fields will be formed constantly along with the time. Defining the hazard position as $P_s$, for example, it can affect 
all agents in its perilous field with radius, defined as $r_s$, in the existence time, defined by $U$. 
If the diffusion velocity and diffusion time for the hazard is $\textbf{v}_s$ and $t_s$, respectively, where $\textbf{v}_s=\{\textbf{v}_{s1},\textbf{v}_{s2}, ... ,\textbf{v}_{sn}\}$, $n \rightarrow \infty$ depicts all possible directions 
for the diffusion, the new hazard point $P_s'$ can be defined as Equation \ref{eq1}:

%¹«Ê½Î´±àÒëͨ¹ý£¨ÒÑÐ޸ģ©
\begin{equation}\label{eq1}
\begin{small}
\begin{array}{l}
P_s^{'} = {P_s} + {t_s} \cdot {{{\textbf{v}}}_{{s}}}\\
%{{\mathop{\rm v}\nolimits}_s}\\
%{\rm{}}
\text{ }\text{ }\text{ }\text{ } = \left\{ {{P_s} + {t_s} \cdot {{{\textbf{v}}}_{{s1}}},{P_s} + {t_s} \cdot {{{\textbf{v}}}_{{s2}}},...,{P_s} + {t_s} \cdot {{{\textbf{v}}}_{{sn}}}} \right\}
\end{array}
\end{small}
\end{equation}

The dangerous range $A_s$ after the diffusion forms a closed area consisting of $P_s$ 
as the source point and all points $P_s'$ as the boundary. 
Then we divide this area into two parts using a line between 1 and ${\frac{{n}}{2}}$, and this area   
can be expressed as the sum of integration of these two parts.

%¹«Ê½Î´±àÒëͨ¹ý£¨ÒÑÐ޸ģ©
\begin{equation}\label{eq2}
{A_s} = \int_1^{\frac{{n}}{2}} {\left( {{P_s} + {t_s} \cdot {{{\textbf{v}}}_{{s}}}  } \right)} 
{d_{{{{\textbf{v}}}_{{s}}} }} - \int_{n}^{\frac{{n}}{2}} 
{\left( {{P_s} + {t_s} \cdot {{{\textbf{v}}}_{{s}}} } \right)} {d_{{{{\textbf{v}}}_{{s}}}  }}
\end{equation}

According to the above description, the dangerous impact on each 
agent is related to the dangerous range of hazard and the distance between 
the hazard and an agent. The farther the 
distance is, the smaller the impact, all points with the same distance from hazard share the same dangerous impact. 
In order to depict this symmetry and attenuation, which is inspired by the work in \cite{1}, 
a Gaussian distribution function is chosen 
to depict this procedure by Equation \ref{eq3}.

\begin{equation} 
\label{eq3} 
\begin{small}
\scalebox{1.0}{$
{{\Gamma }_{s}}\left( P,t \right) =\left \{ 
\begin{array}{lll}\!\! 
 {\!\!\!\!}         \frac{1}{\sqrt{2\pi }\cdot {{r}_{s}}}{{e}^{-\frac{{{\left( P-{{D}_{s}}
   \right)}^{2}}}{2{{r}_{s}}^{2}}}}                         
 {\!\!\!\!}           & \mbox{if} \left\|\left. P-{{D}_{s}} \right\|<{{r}_{s}}\text{ }and\text{ }t\in U \right. \\
           \\
                0
                 &  \mbox{otherwise}\\

\end{array} \right. 
$}
\end{small}
\end{equation}

Here, $\Gamma_{s}(P,t)$ is the strength of danger at the position $P$ produced by 
hazard $s$ at time $t$. $U$ is the duration of hazard $s$.
$D_s$ is the 
intersection position of line $PP_s$ and the hazard area $A_s$($D_s$ can be seen as the hazard position $P_s$ in static hazard situations), and $r_s$ is its influence radius. 
It is noteworthy that danger strength will be 1.0 if position $P$ is within the dangerous range $A_s$.

\section{Emotional Contagion Model Construction}
\label{emotional-contagion}

The emotional contagion model under multi-hazard situations needs to consider 
the panic emotion invoked directly by the hazards, panic propagation among 
individuals, and panic attenuation. The final panic emotion of each 
agent can be obtained by summing up these three components. 

\subsection{Emotional impact from multiple hazards} \label{danger_field} 

In Section \ref{multihazard}, we have defined the perilous field and 
the strength of danger of different hazards. Since the normalized value of the strength of danger is within the range $[0,1]$ , which is the same as the 
property of emotional value \cite{11}, therefore, we adopt the strength of danger, perceived by the agent directly, as the panic value at the current 
position in Equation \ref{eq4}.

\begin{equation}\label{eq4} E_{i}^{h}\left( P,t 
\right)=\sum\limits_{s=1}^{n}{{{\Gamma }_{s}}\left( P,t \right)} 
\end{equation} %notations here 

Here, $E^h_i(P,t)$ represents the panic value of agent $i$ affected by all the hazards $s$
at time $t$ and position $P$, 
where $n$ denotes the total number of hazards.

\subsection{Emotional contagion among individuals} \label{emotional_contagion}
In real life, individuals escaping from the perilous field will carry panic emotion 
and propagate the panic continuously to infect other individuals within a certain distance 
when they are moving. Individuals who perceive this panic may also be affected by them, 
incorporate into their emotions and then pass them out. In addition, emotional contagion 
among different agents are totally different. The extent of emotional transmission among 
agents depend on their personalities, which affect their ability of expression and reception.

In order to depict the above process, we use the emotional contagion model proposed in \cite{17}, which 
incorporate a complex but easy-to-use psychological component into agents to simulate various crowd types. 
one personality model and two thresholds are used in this process. Specifically, OCEAN personality model 
\cite{42} defines a five-dimensional vector $\langle\Psi^O,\Psi^C,\Psi^E,\Psi^A,\Psi^N\rangle$ to characterize the individuals' five kinds of personality:  
openness, conscientiousness, extraversion, agreeableness and neuroticism.   
Each dimension takes a value between -1 and 1. Moreover, personality can also affect the decision of agent in different situations.  
The two thresholds are expressiveness and susceptibility. Expressiveness correlated with extroversion, represents the ability to diffuse emotion.  
Susceptibility represents the minimum value of agent be affected by other agents.  
Taking agent $i$ and agent $j$ as an example, if the emotional value for agent $j$ at certain time is higher than its expressive force threshold, 
it will express the emotion to others. At the same time, if all emotions agent $i$ received exceeds its susceptibility threshlod, 
agent $i$ can be affected by this emotion. The expressiveness threshold for agent $j$ and susceptibility threshlod for agent $i$ are defined as follows:

\begin{equation}\label{eq6}
{e^{{T_j}}} \sim N\left( {0.5 - 0.5\psi _j^E,{{\left( {\left( {0.5 - 0.5\psi _j^E} \right)/10} \right)}^2}} \right)
\end{equation}

\begin{equation}\label{eq7}
sus{T_i}\left( t \right) \sim N\left( {0.5 - 0.5{\varepsilon _j},{{\left( {\left( {0.5 - 0.5{\varepsilon _j}} \right)/10} \right)}^2}} \right)
\end{equation}
Where $N(.,.)$ represents a normal distribution with the former as mean and the later parameter as a standard deviation, 
the empathy value $\varepsilon_i (\varepsilon_i \in [-1,1])$ in Equation \ref{9} for agent $i$ can be described as follows\cite{43} :
\begin{equation}\label{9}
\scalebox{0.85}{$
{\varepsilon _i} = 0.354{\psi ^O} + 0.177{\psi ^C} + 0.135{\psi ^E} + 0.312{\psi ^A} + 0.021{\psi ^N}
$}
\end{equation}

Then for the susceptible agent $i$, all effect caused by all agent $j$ who is expressive and in the perceived range of it at time $t$ can be computed by Equation \ref{eq8}:

\begin{equation}\label{eq8}
E_{i}^c(P,t) = \sum\limits_{t' = t - k + 1}^t {\sum\limits_{j=1}^n {{d_i}\left( {t'} \right)E_{j}^{c'}\left( P_j,t' \right)} }  
\end{equation}

Where $d_i(t')\sim N$(0.1,0.0001) represents the dose values which agent $i$ accepted from agent $j$ at time $t'$, 
$E_j^{c'}\left( {{P_j},t'} \right)$ is the panic emotion of agent $j$ within the perceiving range of agent $i$ at time $t'$. The value of $k$ is set as 10 based on \cite{17}, which means the emotional accumulation of agent $i$ 
at time $t$ is determined by the emotional values in the last 10 consecutive time steps. 

\subsection{Emotion combination}  \label{emotion_combination}  

Based on the documented observations \cite{17}, the panic emotion of individuals will decay 
over time gradually until to the normal state. So we define an emotional attenuation 
function to describe this process, where a parameter $\eta$ is the 
emotional decay rate. For agent $i$ at time step $t$, its new panic can be computed as following:

\begin{equation} \label{eq9}  E_i^d(P,t) = {E_i}({P^{pre}},t - 1) \cdot \eta 
\begin{array}{*{20}{c}} {}&{} \end{array}\eta  \in (0,1]\end{equation}

As mentioned at the beginning of this section, the final panic emotion of 
each agent can be obtained by combining all above three components. 
Considering the Equations \ref{eq4}, \ref{eq8} and \ref{eq9}, the incremental
panic of the agent $i$, who is at the position $P$ and at time $t$, can be 
computed by Equation \ref{eq10}.  With this incremental value, we can obtain 
the panic emotion by Equation \ref{eq11}. It is noteworthy that the 
emotional value $E_i(P,t)$ needs to be normalized after update. 

\begin{equation} \label{eq10} \Delta {E_i}\left( {P,t} \right) = E_i^h\left( 
{P,t} \right) + E_i^c\left( {P,t} \right) - E_i^d\left( {P,t} \right) 
\end{equation} 

\begin{equation}\label{eq11} {E_i}\left( {P,t} \right) = {E_i}\left( 
{{P^{pre}},t - 1} \right) + \Delta {E_i}\left( {P,t} \right) \end{equation}

\section{Emotional Reciprocal Velocity Obstacle}
\label{evacuating}

After the panic of each agent in a multi-hazard environment is computed 
during evacuation, the stressful behaviors of these agents 
affected by the panic emotion can be determined. The location and moving 
direction of an agent are denoted as $P$  and $\mathop V\limits^ \to$, 
respectively.  When the agent has perceived the impact from a hazard $s$ at 
location $P_s$, it will try to follow the \textbf{\emph{stress safety 
direction}} $\mathop {{P_s}P}\limits^ \to$ to escape from the hazard 
instinctively. By contrast, those agents who are not within the impacted area 
of any hazard, will follow their original moving directions. If an 
agent is affected by multiple hazards,  then all the stress safety 
directions of interest will be the result of a weighted sum. So, the stress 
safety direction of an agent in multi-hazard situations can be 
described by Equation \ref{eq12}: 

\begin{equation} \label{eq12} \begin{small}\scalebox{1.0}{$\mathop {V_i^s\left( {P,t} 
\right)}\limits^ \to = \!\! \left \{ \begin{array}{lll} 
\!\!\!          \sum\limits_{s = 0}^{n - 1}{{\Gamma _s}\left( {P,t} \right)} \cdot \mathop{{P_{{s}}}P}\limits^ \to                        
          
           & \mbox{if} \! \left\| \left. P-{{P}_{s}}  \right\|<{{r}_{s}}\text{ }and\text{ }t\in U \right. \\
\!\!\!          \mathop V\limits^ \to 
                 &  \mbox{otherwise}\\

\end{array} \right. 
$}
\end{small}
\end{equation}

\begin{figure}[tb] \begin{centering} 
 \centering
  \includegraphics[width=0.9\columnwidth]{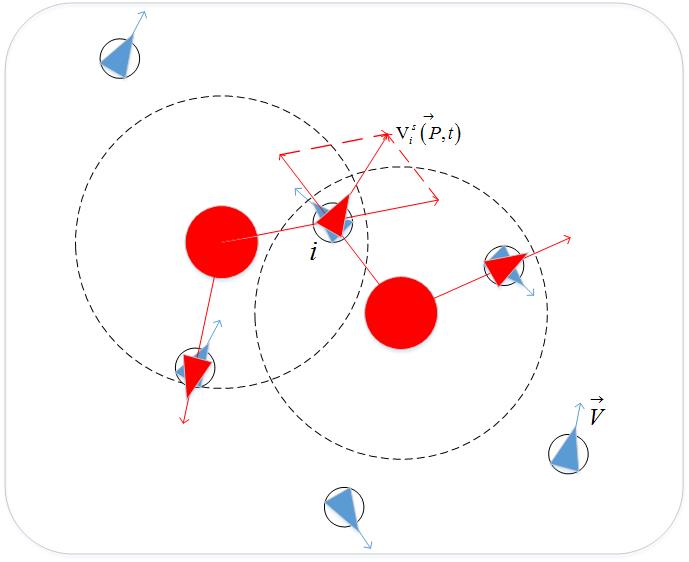}
  \centering
  \caption{Stress safety direction invoked by hazards. Red solid circles 
  represent hazards, dotted circles are the perilous fields 
  of the hazards. The original directions of agents are represented by blue triangles, 
  while the stress safety directions of affected agents are denoted by red triangles.}
  \label{fig:2}
\end{centering}
\end{figure}

Here, $\mathop {V_i^s\left( {P,t} \right)}\limits^ \to$ is defined as the 
\textbf{\emph{safety evacuation direction}} for agent $i$ at the position 
$P$ and time $t$. $U$ is the duration of hazard $s$. Figure \ref{fig:2} 
shows different safety evacuation directions chosen by a group of 
individuals. 

Besides the direct emotional impact from hazards, the contagious panic 
emotion received from its neighbors may also alter agents' original moving 
directions. As mentioned in \cite{2}, we assume the probability of agent $i$ 
following its original direction is $p_i$ and the probability $1-p_i$ to 
follow the others' directions. Thus, the new direction can be defined as the 
addition of these two direction vectors. In this paper, the probability 
$p_i$ is equal to the panic value ${E_i}(P,t)$ of agent $i$. The updated 
moving direction of agent $i$ at time $t$ is defined as below:

\begin{equation} 
\label{eq13}
\scalebox{0.8}{$ 
\mathop {V_i^c\left( {P,t} 
\right)}\limits^ \to   = {{E_i}(P,t)\mathop {V_i^s(P,t)}\limits^ \to   + 
\left( {1 - {E_i}(P,t)} \right)\sum\limits_{j\in R(i)} {\mathop 
{V_j^c({P_j},t)}\limits^ \to  } }
$} 
\end{equation}

Here, $\mathop {V_i^c\left( {P,t} \right)}\limits^ \to$ represents the moving 
direction of agent $i$ who is at the position $P$ at time $t$. 
$\sum\limits_{j\in R(i)} {\mathop {V_j^c\left( {{P_j},t} \right)}\limits^ 
\to }$ is the combined moving directions of those agents who are in the 
emotional perception range of agent $i$. 
$R(i)$ denotes neighboring agents within the perception range of agent i.
When the agent is going to change its direction, we 
assume the magnitude of its velocity will remain. In other words, 
the velocity module of the agent at that time should be $\textbf{V}_{{i}}^{{c}}$.  

\begin{figure*}[htb]
\begin{centering}
 \centering
  \includegraphics[width=0.8\linewidth]{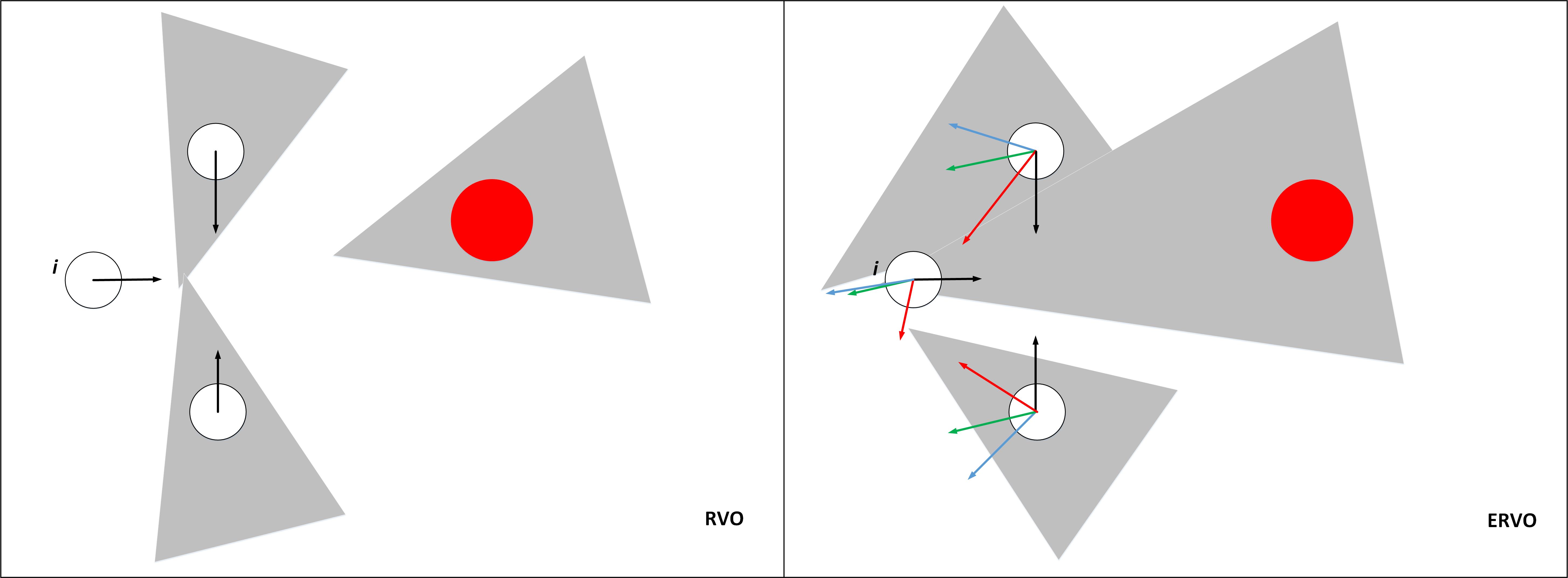}
  \centering
  \caption{The collision area computed by the traditional RVO model and our ERVO model for agent $i$. 
  Grey triangle areas around white circles and red solid circles represent the collision areas caused by agents and hazards, respectively. 
  The black, blue, green and red arrows are separately the original direction, 
  stress safety direction, emotional contagion direction and final direction 
  of one agent. The emotional contagion direction of an agent is determined 
  by combining its safety stress direction with those of its neighbors. The 
  final direction is determined by combining its original direction and emotional 
  contagion direction.} 
\label{fig:3} 
\end{centering} \end{figure*}

In Equation \ref{eq13}, the moving direction of agent $i$ is only influenced 
by panic emotion. However, in the actual crowd movement, the final direction 
of an agent is also influenced by its 
planned targets and other neighboring moving agents. In other words, the local obstacle avoidance and global 
path planning for agents also need to be considered. The RVO model \cite{8} 
is an efficient and safe multi-object automatic navigation algorithm. 
However, during the obstacle avoidance, the RVO model focuses on the 
position and velocity of the current agent and other agents (refer to Equation \ref{eq14}), but does not take into account the emotional impact on 
speed selection invoked by surrounding obstacles and existing hazards. In 
Equation \ref{eq14}, $RVO_j^i({{{\textbf{V}}}_{{j}}},{{{\textbf{V}}}_{{i}}},\alpha _j^i)$ is the collision area for agent $i$ caused by agent $j$ 
(illustrated in the grey area around the white circle of Figure \ref{fig:3} (RVO)), which means that agent $i$ and agent $j$ will collide with each other once the velocity of agent $i$ fall into this area. $\textbf{V}_i$ and $\textbf{V}_j$ represent the velocity for agent $i$ 
and agent (or hazard) $j$. $\alpha^i_j$ is the effort chosen by agent $i$ to 
avoid the collision with agent (or hazard) $j$, which is implicitly assumed to ${\frac{1}{2}}$ in the original RVO model. 
For more details of the RVO model, please refer to \cite{8}.

\begin{equation}\label{eq14} 
\scalebox{0.8}{$
\begin{array}{l} 
RVO_j^i({{{\textbf{V}}}_{{j}}},{{{\textbf{V}}}_{{i}}},\alpha _j^i) = \{ 
{{\textbf{V}}}_{{i}}^{{'}}|\frac{1}{{\alpha _j^i}}{{\textbf{V}}}_{{i}}^{{'}}{ {                               
}} + (1 - \frac{1}{{\alpha _j^i}}){{{\textbf{V}}}_{{i}}} \in VO_j^i({{{\textbf{V}}}_{{j}}})\} 
\end{array}
$}
\end{equation}

Inspired by the RVO model, we propose a new ERVO model by integrating 
emotional contagion into crowd movement planning. This new model 
constructs a new collision area (shown by the grey triangle areas in Figure \ref{fig:3} (ERVO)) by considering 
the current velocity $\textbf{V}_i$ and the updated velocity $\textbf{V}_i^c$ of the agent, and 
also the velocity $\textbf{V}_j$ as described in Equation \ref{eq15}. The 
effort made by agent $i$ to avoid collision with agent (or hazard) $j$ is 
defined in Equation \ref{eq16}. 

\begin{equation}\label{eq15} 
\begin{array}{l} 
ERVO_j^i({{{\textbf{V}}}_{{j}}},{{{\textbf{V}}}_{{i}}},{{\textbf{V}}}_{{i}}^{{c}},\alpha _j^i) = \{ 
\textbf{V}_i'|\frac{1}{{\alpha _j^i}}({{\textbf{V}}}_{{i}}^{{'}} + {{\textbf{V}}}_{{i}}^{{c}})\\ { {                                    
}} + (1 - \frac{1}{{\alpha _j^i}}){{{\textbf{V}}}_{{i}}} \in VO_j^i({{{\textbf{V}}}_{{j}}})\} 
\end{array} 
\end{equation} 

\begin{equation} \label{eq16} \alpha _j^i = \frac{{{E_j}(P,t)}}{{{E_i}(P,t) 
+ {E_j}(P,t)}} \end{equation} 

During the crowd simulation, for agent $i$, if $\textbf{V}_i$ is outside of the 
emotional reciprocal velocity obstacle of agent (or hazard)$j$, both of them 
will never collide. The ERVO model can be used to navigate a large number of 
agents in a complex multi-hazard scenario. For each agent $i$ in the 
scene, it has a current position $P$, a current velocity $\textbf{V}_i$, an updated 
velocity ${{\textbf{V}}}_{{i}}^{{c}}$, a current panic emotion ${E_i}(P,t)$, and a 
goal location $G_i$. For a hazard $s$, it has position $P_s$ and duration $t$. For obstacle $o$, it has current position $P_o$ and velocity 
$\textbf{V}_o$. Static obstacles have zero velocity in particular. In our experiments, 
we choose a small time step $\Delta t$ to simulate crowd behaviors. Within 
this time step, we select a new velocity for each object independently and 
update its position according to the surrounding environment until all of 
the agents have reached the safe area or their goals.

\section{Experiment Results} 
\label{experiment}

We run a diverse set of crowd simulations in multi-hazard situations, 
all experiments are realized by using C++ in the Visual Studio and Unity 3D platform.  
Our experiment results show that our method can soundly generate realistic 
movement as well as panic emotion dynamics in a crowd. In Section 
\ref{7.1}, we simulate crowd behaviors in four different outdoor multi-hazard scenes. 
In Section \ref{7.2}, we analyze the importance of our emotional contagion mechanism and different 
influence in different scenarios.  
Then the emotional contagion model is proved more suitable for our multi-hazard situations in Section \ref{7.3}. 
Furthermore, we validate the realism of our simulation results by comparing 
them with the crowd movement in real world in Section \ref{7.4} and the 
effectiveness of our method in different virtual environments in Section 
\ref{7.5}. 
  
\subsection{Crowd simulation under different multi-hazard scenarios} 
\label{7.1} 

As discussed before, different hazard types have various effects on crowd 
movement. We simulate emergency behaviors in a crowd with the following 
two-hazard situations: (1) persistent hazards occur at the same time; (2) 
transient hazards occur at the same time; (3) persistent hazards occur at different moments; (4) transient hazards occur at different moments. All 
simulations run in open field, and each simulation involves forty agents. The persistent hazards and transient hazards are represented by fire and explosion, respectively. The time step is set to 
0.25s, other parameters in our system are set experimentally: 
the influence radius ${{r}_{s}}$ = 10 m, 
emotional decay parameter $\eta$ = 0.01, the personality parameters $\Psi^O, \Psi^C, \Psi^E, \Psi^A, \Psi^N$ are set to the random number between -1 and 1 for simplify to depict different agents,    
and the perceived scope is set to 4 for all agents. 

\begin{figure*}[!htbp] 

{\centerline{\includegraphics[width=6in]{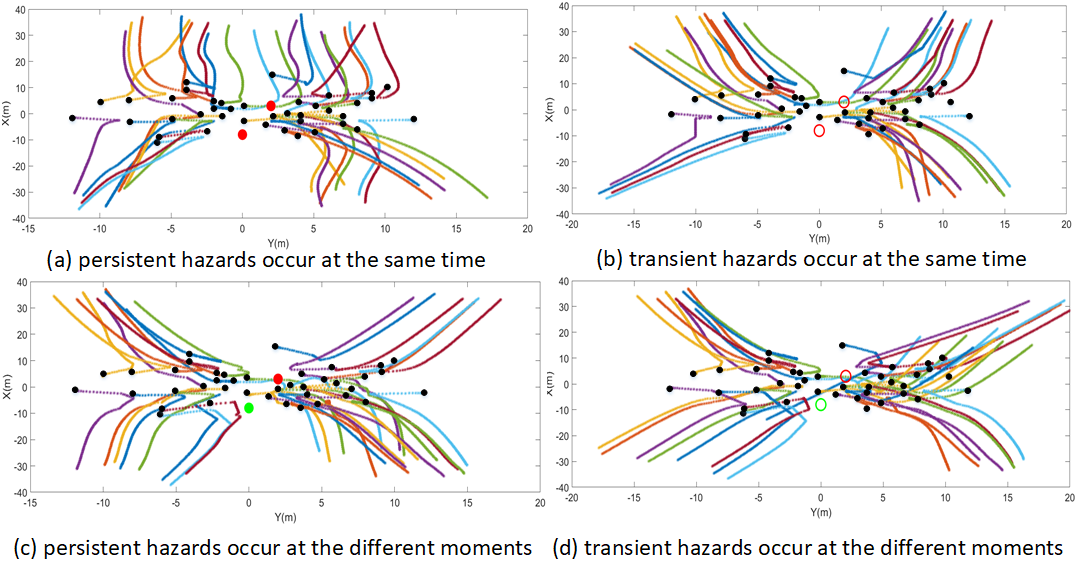}}} 

\caption{\label{fig:4}
Movement trajectories of forty agents in different types of hazard scenarios. In each scenario, black points represent the initial positions of all agents, the lines drawn by different colors are used to depict 
different paths of agents, while trajectories for the same agent use the same color in different conditions. 
In addition, 
the red solid and hollow circles represent persistent and transient hazard positions, respectively. 
While the green one represent the positions of the second hazards in the concurrent conditions.
} \end{figure*} 

Path flow maps for all agents are used to depict the crowd movement differences among this four conditions. 
As illustrated in Figure \ref{fig:4}, the black points are the original positions of all agents, lines of different colors are used to depict different paths of agents, while trace 
flows for the same agent indicated by the same color in four conditions, the red solid and hollow circles 
represent persistent and transient hazard positions in our scenarios, respectively. While the   
second hazards occur in the concurrent conditions draw by green.

If two hazards occur at the same time as shown in Figure \ref{fig:4}(a), Figure \ref{fig:4}(b). 
Agents around these two hazards will change their routes to be distant far away 
from them. When compared with the transient condition, based on the persistent effect from hazards, more emotional 
contagion lead to jittery for many paths of agents (shown in Figure \ref{fig:4}(a)), while trajectory for the agents in transient 
conditions are smoother owing to the disappear of hazards in this scenario (shown in Figure \ref{fig:4}(b)).

If two hazards occur at different moments, as shown in Figure \ref{fig:4}(c), Figure \ref{fig:4}(d). When the first hazard occurs, 
agents in the perilous field of this hazard will change their movement direction far away from it, while other agents keep the original movement. 
When the second hazard occurs, if the first one does not disappear, agents 
will escape away from both of the two hazards (shown in Figure \ref{fig:4}(c)). By contrast, some agents' path may move to or pass through 
the area where the first hazard disappeared (shown in Figure \ref{fig:4}(d)).

\begin{figure}[htb]
\begin{centering}
 \centering
  \includegraphics[width=0.8\linewidth]{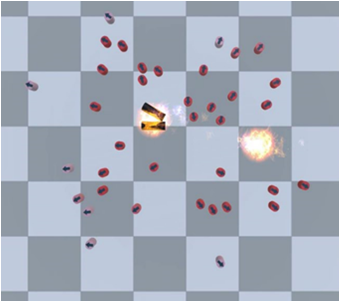}
  \centering
  \caption{One snapshot in the condition of persistent hazards occur at the same time, where the cylinders 
  are used to depict agents, different colors represent different panic values of them, the darker the color is, the larger its panic value.} 
\label{fig:5} 
\end{centering} \end{figure}

In addition to that, the panic emotion changes of agents are also important during this procedure. Figure \ref{fig:5} illustrate a snapshot in  
the condition of persistent hazards occur at the same time, where we use a cylinder to represent an agent and visualize 
its panic value using different colors. Despite those two dead agents drawn by the black cylinders, the white, light red, 
red, dark red and red black are used to represent ${E_i} = 
0$, ${E_i} \in (0,0.3]$, ${E_i} \in (0.3,0.5]$, ${E_i} \in (0.5,0.7]$ and 
${E_i} \in (0.7,1.0]$, respectively. The larger the panic value is, the darker its color. 
For more dynamic simulation details in different multi-hazard conditions, we refer readers to our supplemental video.

\subsection{Analysis of emotional contagion}\label{7.2}

\begin{figure}[!htbp]   
 \centering
 \subfigure[without emotional contagion model]{
  \includegraphics[width=1.0\columnwidth]{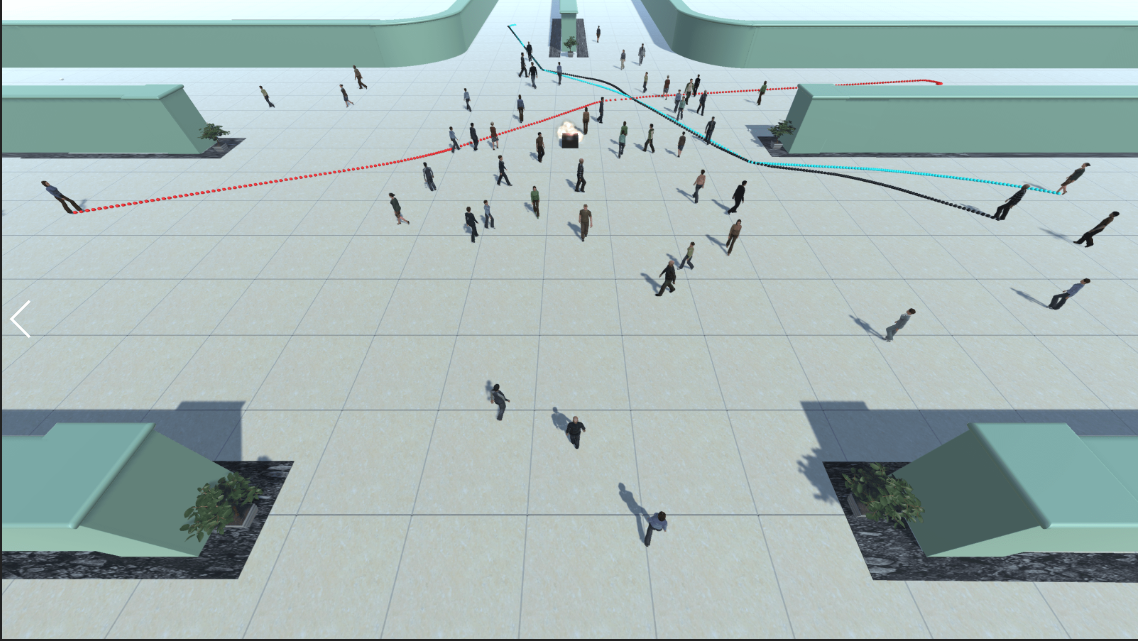}
 }
\subfigure[with emotional contagion model]{ 
  \includegraphics[width=1.0\columnwidth]{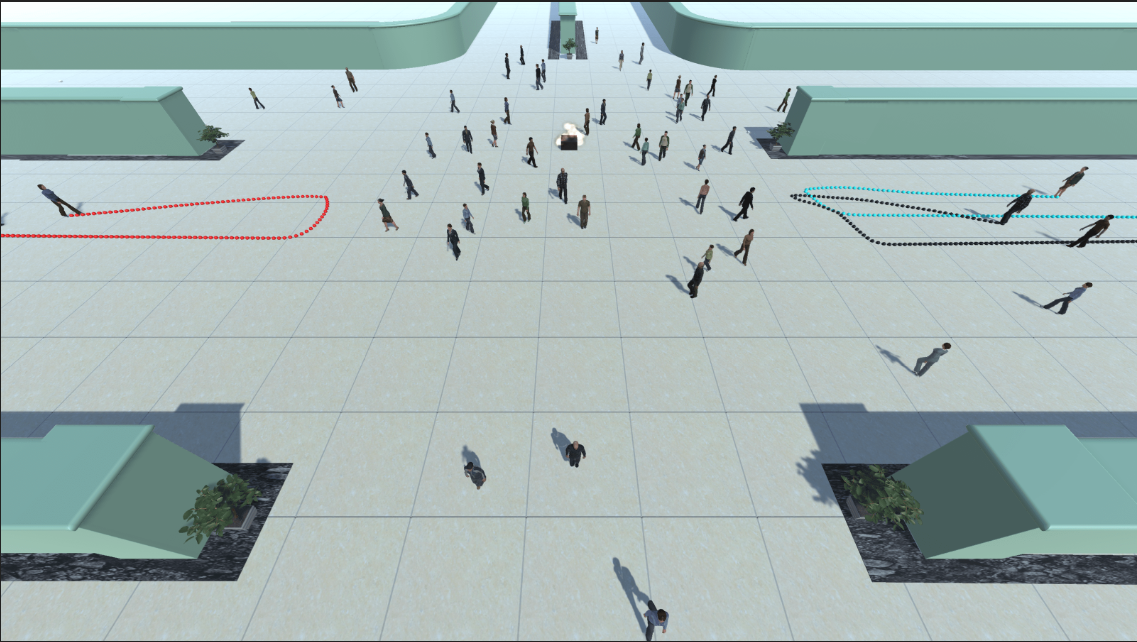}
 } 
 \caption{The comparison of crowd movements with and without emotional contagion.}
 \label{fig:6}
 \end{figure}

In order to validate the effectiveness of emotional contagion in our method, 
we run crowd simulations in a scene with and without this mechanism, 
respectively. Figure \ref{fig:6} shows the moving trajectories of three 
selected agents in the situation with one transient hazard. Agents with 
emotional contagion will adjust their moving directions to escape away from 
the hazard even when they have not reached the nearby region of the hazard. 
In contrast, agents without emotional contagion will keep moving along the 
original planned directions. The trajectory of one agent is illustrated by 
one colorful line. From these results, we can infer that the crowd movement 
in a hazard environment is affected by the panic emotion significantly.  

 \begin{figure}[!htbp] 
 \centering
 \subfigure[panic changes caused by persistent hazards]{
 \begin{minipage}{0.45\textwidth}
  \includegraphics[width=1.0\columnwidth]{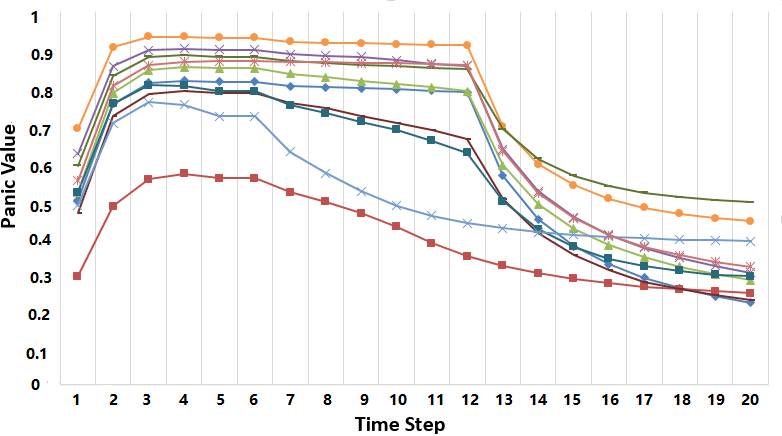}
  \end{minipage}
 }
\subfigure[panic changes caused by transient hazards]{ 
 \begin{minipage}{0.45\textwidth}   
  \includegraphics[width=1.0\columnwidth]{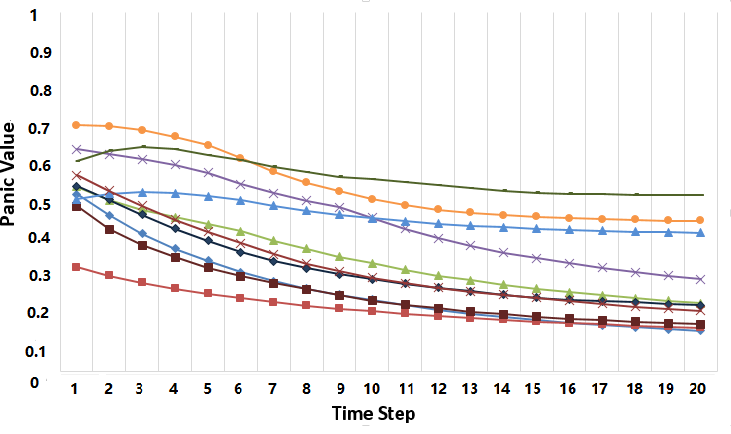}
 \end{minipage}
 } 
 \caption{The panic emotion changes in a crowd in different situations. 
The simulation contains 15 agents and each colored line represents the 
panic emotion of one agent in the scene. } 
 \label{fig:7}
 \end{figure}

In the previous section, we have discussed the effect of emotional propagation 
on crowd movement qualitatively. Here we mainly focus on the change of panic 
emotion of each agent during the crowd evacuation, especially when 
persistent/transient hazards occur at the same time. From Figure 
\ref{fig:7}, we can see that the panic emotion value will increase to the 
maximum when a persistent hazard happens. The reason is that although the 
agent is moving far away from the hazard, the agent is still in the perilous 
field and the panic value is accumulated. When agents are out of the 
perilous field, their panic values will decay and reach to a similar low 
level due to the effect of emotion contagion. For a transient hazard, the 
panic emotion will reach to the maximum immediately when the hazard occurs, 
then it will decrease gradually.

\subsection{Comparisons with another emotional contagion model}\label{7.3}

\begin{figure}[!htbp] 
 \centering
 \subfigure[panic emotional interval distribution over time in model \cite{44}]{
 \begin{minipage}{0.45\textwidth}
  \includegraphics[width=1.0\columnwidth]{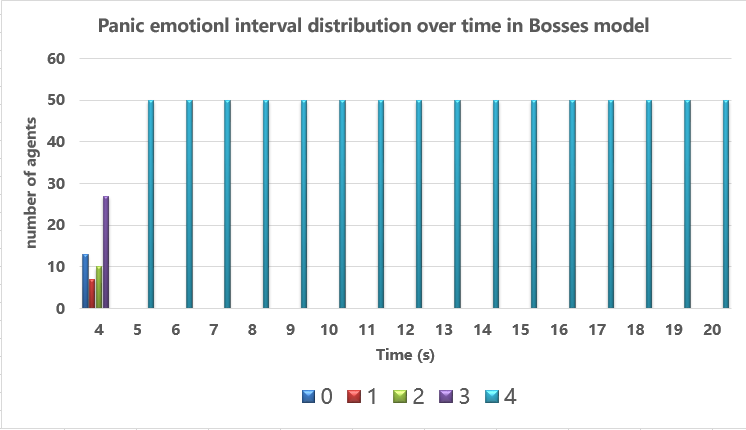}
  \end{minipage}
 }
\subfigure[panic emotional interval distribution over time in our model]{ 
 \begin{minipage}{0.45\textwidth}   
  \includegraphics[width=1.0\columnwidth]{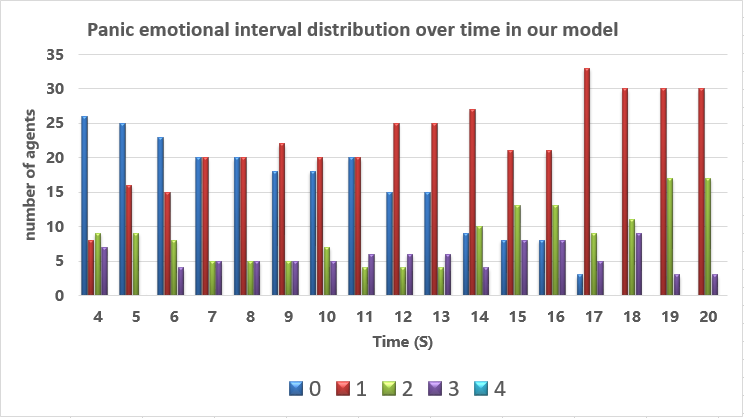}
 \end{minipage}
 } 
 \caption{The agent numbers in different panic emotional interval during the evacuation. 
five panic emotion levels depicted by 0-4 with different colors, the higher this value, the higher the panic emotion.} 
 \label{fig:8}
 \end{figure}

\begin{figure}[!htbp] 
 \centering
 \subfigure[with our emotional contagion model]{
  \includegraphics[width=1.0\columnwidth]{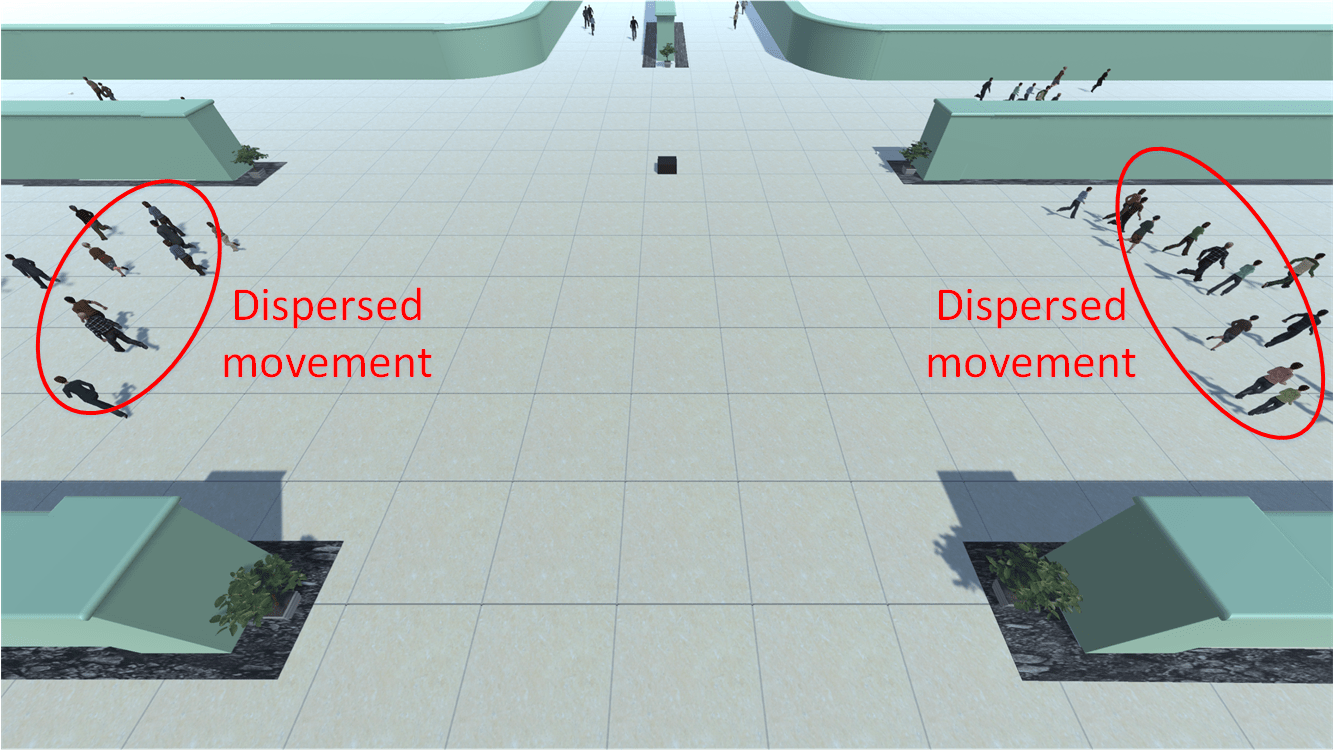}
 }
\subfigure[with emotional contagion model proposed in \cite{44}]{ 
  \includegraphics[width=1.0\columnwidth]{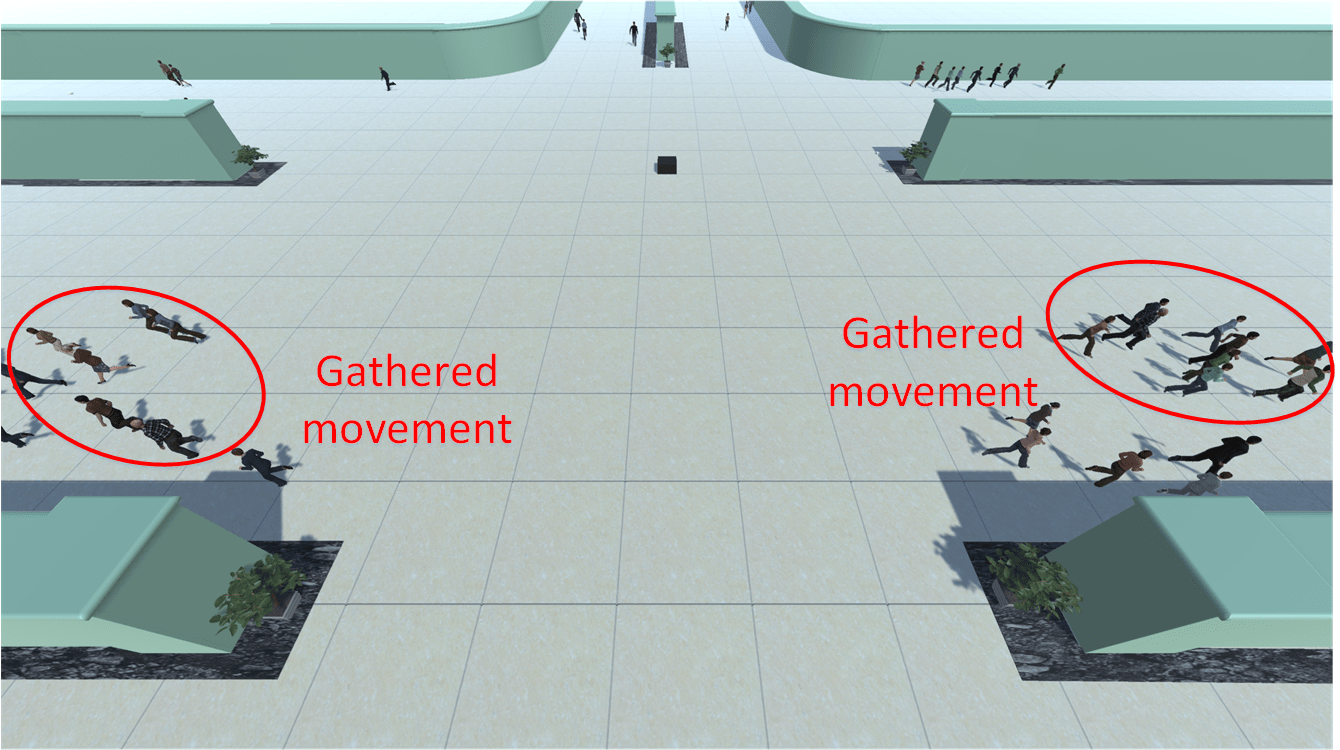}
 } 
 \caption{The comparison of crowd movements with another emotional contagion model.}
 \label{fig:9}
 \end{figure} 

In order to validate the effectiveness of our emotional contagion model among agents, we 
compare our simulation results to an agent-based emotional contagion model proposed in 
\cite{44}. Same personality and original state are chosen for fifty agents in this two models,  
then the overall difference caused by emotional contagion can be caught. 
After bomb occurs, agents may have different panic 
emotions and movements in different time. Panic emotion of all  
agents and movements simulation results can be shown in Figure \ref{fig:8}, Figure \ref{fig:9}.

In Figure \ref{fig:8}, the number distribution of agents panic emotion are illustrrated by five levels defined in Section \ref{7.1}, 
Where 0 as the lowest panic emotion values 0 and level 5 represents the highest panic emotion values from 0.7 to 1.0. 
We choose the explosion time at 4s as the start time, which can be seen that all agents have the high panic emotion 
almost the whole evacuation process when used emotional contagion model mentioned in \cite{44}, but in our model, 
the number of lower emotion levels decrease first and increase as following, the higher level ones reverses. 
The reason of this phenomenon is that \cite{44} considers all agents in the whole scenes once a hazard occurs, and 
does not take emotion decay into account. While in our emotional contagion model, each agent have a perception 
range as well as expressiveness and susceptibility to accept emotional contagion from others, and their panic 
emotion change along with the movement. 

In addition, the simulation results in this two conditions shown in Figure \ref{fig:9}. The movement of agents 
after explosion in our model are more dispersed as labeled by red ellipses, while in another model, all agents 
behave towards an aggregation states. With considering the different emotion changes in this two models, lower 
panic emotion lead to a more independent movement direction (shown in Figure \ref{fig:9}(a)) instead of gathered 
movement based on stronger emotional contagion(shown in Figure \ref{fig:9}(b)). In real world, panic 
emotions will decrease when the crowd are away from hazards. From the results of these two different models, where 
be seen that our emotional contagion model is more realistic and suitable to simulate the crowd movement in the multi-hazard situations.

\subsection{Comparisons with real-world crowd behaviors}\label{7.4}

\begin{figure*}[htbp] 
\subfigure[Recorded video data
(ground-truth)]{\centerline{\includegraphics[width=6.3in]{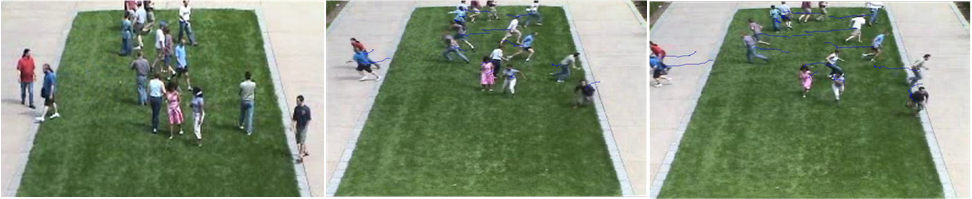}}} 
\subfigure[Our simulation result (corresponding to 
(a))]{\centerline{\includegraphics[width=6.3in]{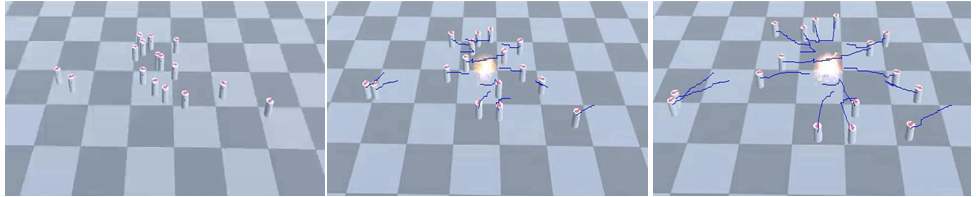}}} 
\subfigure[Recorded video data 
(ground-truth)]{\centerline{\includegraphics[width=6.3in]{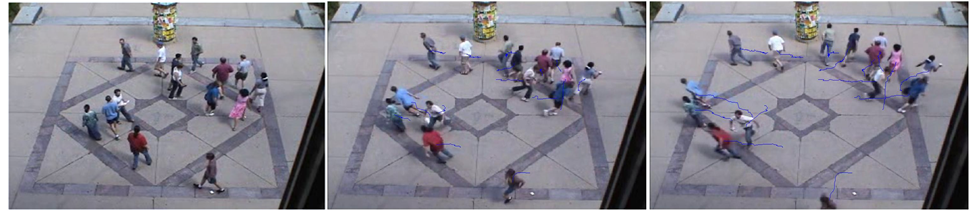}}} 
\subfigure[Our simulation result (corresponding to 
(c))]{\centerline{\includegraphics[width=6.3in]{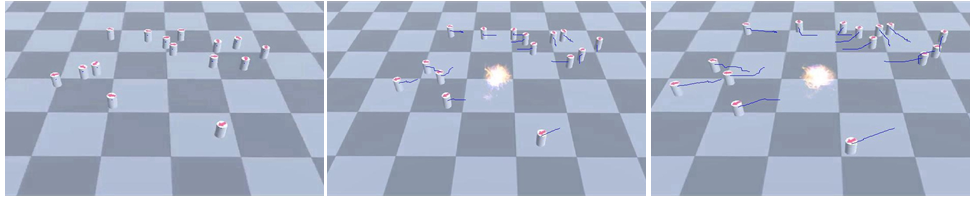}}} 
\caption{\label{fig:10}Snapshots of ground-truth crowd evacuations on the 
outdoor ground and our corresponding simulation results. Three images from 
left to right is: initial random status, at the beginning of the evacuation, 
one moment in the evacuation. The movement trajectories for all agents 
are drawn by the blue lines.} 
\end{figure*}

In order to validate our approach, we also compare the simulation results with real-world crowd evacuation video. 
Two crowd evacuation video are chosen in this part, first one is chosen from the public available dataset of normal crowd videos from University of Minnesota (UMN) \cite{bird2006real}, which is designed 
to test the abnormal detection method originally. In this scene, movement details are used to verify the similarity between 
real-world crowd behaviors and our simulation results. Although no pre-defined goals are set in advance, agents can still be driven 
to escape in a realistic way by our method.Illustrated in Figure \ref{fig:10}, 
three images in each row are the crowd movement states at initial random conditions, at the beginning of the evacuation, one moment after the hazard occurs. The trajectories of agents are shown 
by the blue lines. From the trajectories we can find that movement trends of our simulation results are similar with that in real scenes. 
Furthermore, we also compare the trajectory length and the maximum speed of each agent from the moment when hazard occurs to the end of simulation,  
as shown in Table \ref{tab:1} (comparison with the grassland scene) and Table \ref{tab:2} (comparison with the square scene), 
where can be seen that our mean trajectory length and mean maximum speed are close to the true video data. 
Thus we can see both the overall movement trend of the crowd and individuals' movement details in the crowd are similar to those in 
the recorded real-world crowd video.
More animation comparison details can be found in our supplemental video.

\begin{figure}[htbp]
\begin{centering}
 \centering
 \subfigure[True video data (911 terrorist attacks)]{
  \includegraphics[width=0.8\columnwidth]{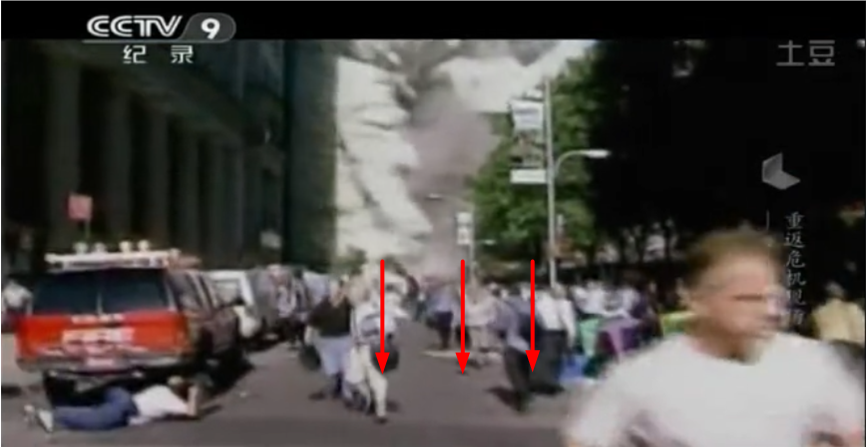}
 }
\subfigure[Simulation result]{ 
  \includegraphics[width=0.8\columnwidth]{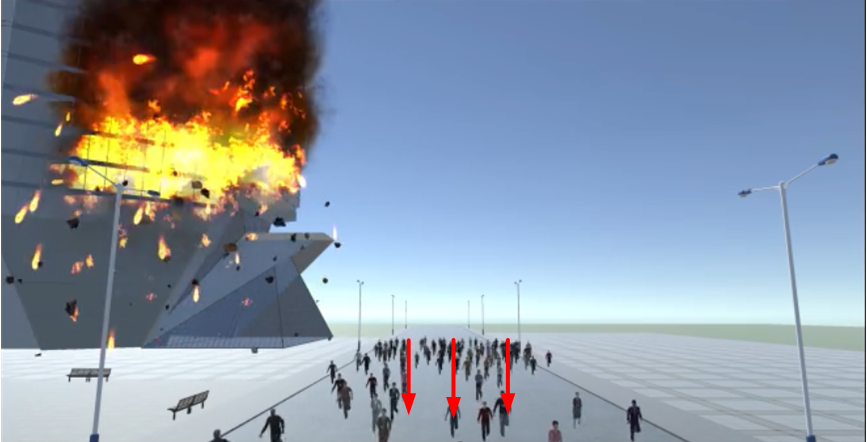}
 } 
  \centering
  \caption{Snapshots of the true video and our simulation result of 911 scenario, where the red arrows represent the movement trends of crowd.} 
\label{fig:11} 
\end{centering} \end{figure}

The second one is the 911 terrorist attacks with two explosion,   
while when considering the camera shaking and crowd occlusion, movement details of agents cannot be obtained accurately, 
thus this scene is mainly used to verify the similarity of group movement trends between our simulation result and true situation. 
In this circumstance, two bombs occurred concurrent on the building, and all agents straight forward in the whole procedure.  As shown in Figure \ref{fig:11}, the crowd movement directions are indicated by red arrows, more details can be 
found in our supplemental video.

\newcommand{\tabincell}[2]{\begin{tabular}{@{}#1@{}}#2\end{tabular}}
\begin{table}
\tiny
\centering
\caption{The comparison between our simulation result and real grassland scene}
\setlength{\tabcolsep}{1mm}{
\begin{tabular}{ccccc}
\hline
\multicolumn{1}{l}{ Agent ID} & \begin{tabular}[c]{@{}c@{}} Trajectory length\\in simulation result\\(pixel)\end{tabular} & \begin{tabular}[c]{@{}c@{}}Trajectory length\\in real video\\(pixel)\end{tabular} & \begin{tabular}[c]{@{}c@{}}Maximum speed\\in simulation result\\(pixel/frame)\end{tabular} & \begin{tabular}[c]{@{}c@{}}Maximum speed\\in real video\\(pixel/frame)\end{tabular} \\ \hline
1                            & 111.8814                                                                                     & 94.1773                                                                            & 6.0828                                                                                   & 8.0156                                                                        \\ 
2                            & 121.1261                                                                                     & 121.4284                                                                             & 5.5902                                                                                   & 8.0000                                                                        \\ 
3                            & 117.4920                                                                                     & 155.8467                                                                             & 5.8310                                                                                   & 9.0000                                                                        \\ 
4                            & 106.193                                                                                     & 89.5567                                                                             & 5.4083                                                                                   & 9.8234                                                                        \\ 
5                            & 94.7727                                                                                    & 50.7417                                                                             & 5.3852                                                                                   & 6.2560                                                                         \\ 
6                            & 86.9983                                                                                     & 32.9509                                                                             & 8.0623                                                                                   & 6.5765                                                                        \\ 
7                            & 83.7936                                                                                     & 60.6989                                                                             & 5.0000                                                                                   & 3.6056                                                                        \\
8                            & 110.8647                                                                                     & 110.0192                                                                             & 6.5192                                                                                   & 11.5109                                                                       \\ 
9                            & 57.6991                                                                                     & 61.0370                                                                             & 5.5227                                                                                   & 9.8489                                                                        \\ 
10                           & 45.4759                                                                                     & 42.4537                                                                             & 3.6401                                                                                   & 7.5664                                                                        \\ 
11                           & 35.5255                                                                                      & 39.0000                                                                           & 3.5355                                                                                   & 8.5586                                                                        \\ 
12                           & 73.6405                                                                                     & 42.2433                                                                             & 4.7170                                                                                    & 5.3852                                                                        \\ 
13                           & 84.6207                                                                                     & 81.6077                                                                             & 3.5355                                                                                   & 9.0000                                                                         \\ 
14                           & 90.2147                                                                                     & 98.1849                                                                             & 5.0000                                                                                   & 10.0000                                                                       \\ 
15                           & 96.3887                                                                                     & 116.4673                                                                             & 7.5000                                                                                   & 10.0000                                                                       \\ 
16                           & 113.0668                                                                                    & 128.4863                                                                             & 4.5277                                                                                   & 13.5370                                                                        \\ \hline
mean                         & 95.3169                                                                                     & 88.3267                                                                             & 5.3661                                                                                       & 8.5405                                                                              \\ \hline       

\label{tab:1}
\end{tabular}}
\end{table}

\begin{table}
\tiny
\centering
\caption{The comparison between our simulation result and real square scene}

\setlength{\tabcolsep}{1mm}{
\begin{tabular}{ccccc}
\hline
\multicolumn{1}{l}{Agent ID} & \begin{tabular}[c]{@{}c@{}}Trajectory length \\in simulation result\\(pixel)\end{tabular} & \begin{tabular}[c]{@{}c@{}}Trajectory length \\ in real video\\(pixel)\end{tabular} & \begin{tabular}[c]{@{}c@{}}Maximum speed \\ in simulation result\\(pixel/frame)\end{tabular} & \begin{tabular}[c]{@{}c@{}}Maximum speed \\ in real video\\(pixel/frame)\end{tabular} \\ \hline
1                            & 84.9795                                                                                     & 93.1883                                                                             & 6.8007                                                                                   & 5.5902                                                                        \\ 
2                            & 36.1059                                                                                     & 49.7452                                                                             & 4.6098                                                                                   & 4.5277                                                                        \\ 
3                            & 32.0000                                                                                    & 56.9671                                                                            & 4.0000                                                                                   & 3.5355                                                                        \\ 
4                            & 39.0000                                                                                     & 70.1650                                                                             & 3.6056                                                                                   & 7.1589                                                                        \\ 
5                            & 65.1402                                                                                     & 121.4701                                                                             & 5.4083                                                                                   & 9.7082                                                                        \\ 
6                            & 37.3852                                                                                     & 64.8521                                                                             & 3.2016                                                                                   & 6.5765                                                                        \\ 
7                            & 32.3006                                                                                    & 66.2214                                                                             & 3.2016                                                                                   & 3.6056                                                                        \\ 
8                            & 85.3429                                                                                     & 135.1380                                                                             & 5.5902                                                                                   & 9.5525                                                                        \\ 
9                            & 86.2274                                                                                     & 80.3388                                                                             & 6.2650                                                                                   & 6.1033                                                                        \\ 
10                           & 38.7559                                                                                     & 88.4081                                                                             & 5.0249                                                                                   & 6.0208                                                                        \\ 
11                           & 136.2666                                                                                     & 83.5291                                                                             & 12.2577                                                                                  & 6.5192                                                                        \\ 
12                           & 109.6164                                                                                     & 120.4088                                                                             & 7.0711                                                                                   & 9.0139                                                                        \\ 
13                           & 124.5751                                                                                    & 69.5775                                                                             & 13.0096                                                                                  & 13.2004                                                                       \\ 
14                           & 159.8465                                                                                     & 152.3059                                                                             & 9.1788                                                                                   & 10.5000                                                                       \\ 
15                           & 176.6732                                                                   & 191.3627                                                                             & 12.0934                                                                                  & 8.0156                                                                        \\ \hline
mean                         & 82.9477
                                                                                     & 96.2452                                                                             & 6.7546                                                                                       & 7.3086                                                                            \\ \hline        
\label{tab:2}
\end{tabular}}
\end{table}

\subsection{Applications in different scenarios}\label{7.5} 

\begin{figure*}[!htbp]
\centering
\begin{tabular}{cc}
 \includegraphics[scale=0.55]{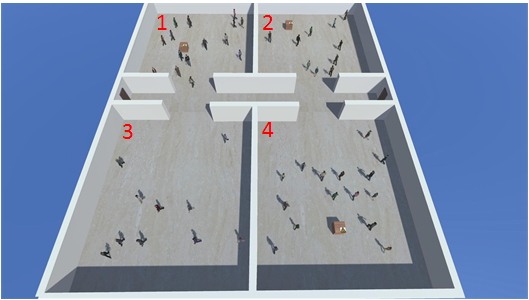} &  \includegraphics[scale=0.55]{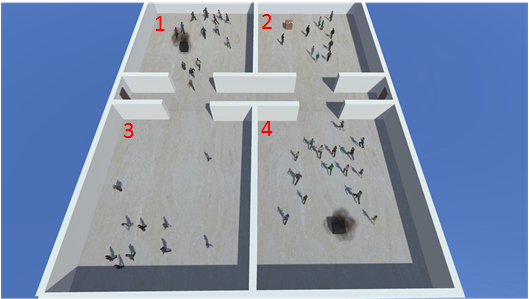} \\
 a ( 1st frame ) & b ( 8th frame ) \\
 \includegraphics[scale=0.55]{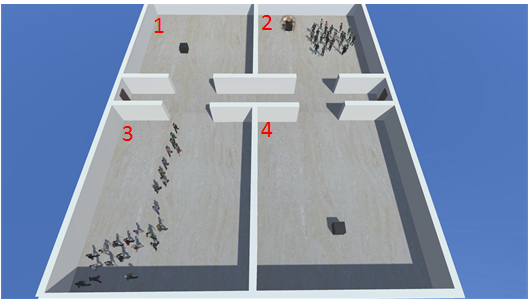} &  \includegraphics[scale=0.55]{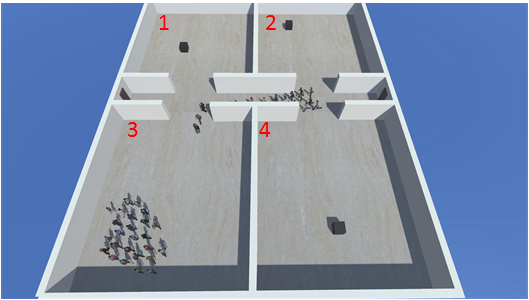} \\
 c ( 64th frame ) & d ( 120th frame ) 
\end{tabular} \caption{Evacuation simulation result by our approach in an 
office building} \label{fig:12} \end{figure*}

\begin{figure*}[!htbp]
\centering
\begin{tabular}{cc}
 \includegraphics[scale=0.55]{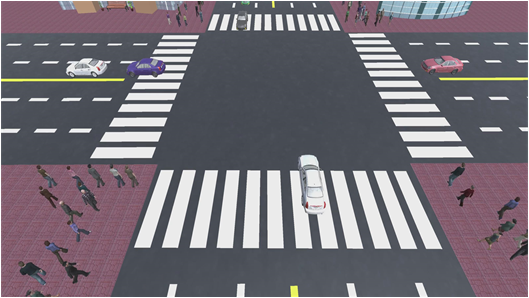} &  \includegraphics[scale=0.55]{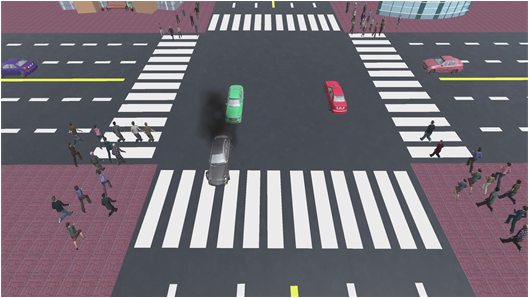} \\
 a ( 1st frame ) & b ( 16th frame )  \\
 \includegraphics[scale=0.55]{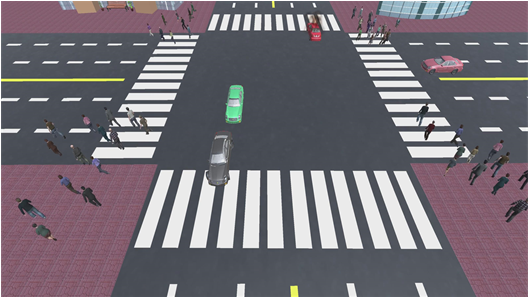} &  \includegraphics[scale=0.55]{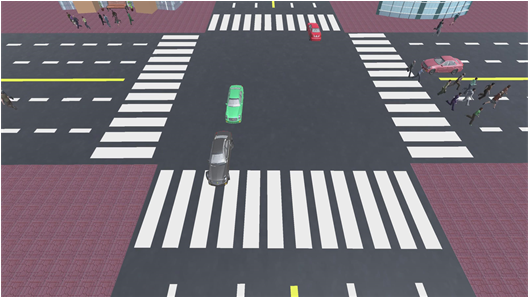} \\
 c ( 24th frame ) & d ( 60th frame ) 
\end{tabular} \caption{Evacuation simulation result by our approach in a
crossroad.} \label{fig:13} \end{figure*}

We apply our method to simulate crowd evacuation simulations in office 
building ( Figure \ref{fig:12}) and crossroads ( Figure \ref{fig:13} ) with 
multiple hazards to check the effectiveness of our method. 

In an office building, we numbered its four rooms as 1, 2, 3, 4 from left to 
right and up to bottom. The corridor in the middle connects all these rooms 
together and there are no exits on both sides. At the beginning, 50 agents 
located in different rooms move randomly in Figure \ref{fig:12}(a) . At the 
8$th$ frame, there are two bomb explosions in room 1 and room 4 at the same 
time in Figure \ref{fig:12}(b).  At the 64$th$ frame,  there is a fire in 
room 2 in Figure \ref{fig:12}(c). From the simulation results, we observe 
the following:  when bomb explosions occur, in order to avoid the danger, 
agents in the rooms begin to move to room 2 and room 3, 
respectively. When room 2 is on fire, the agents in or aiming to room 2 try 
to escape. At last, all of the agents move to the safe room 3 in Figure 
\ref{fig:12}(d). 

The crossroad scene contains 50 pedestrians and two non-current car bombs. 
When the simulation starts, agents cross the road freely in Figure 
\ref{fig:13}(a). At the 16$th$ frame, one black car bomb explodes in Figure 
\ref{fig:13}(b) and another red car bomb explodes at the 24$th$ frame as 
shown in Figure \ref{fig:13}(c). When the first car bomb occurs, the agents 
nearby evacuate immediately. Some agents affected by their neighbors move 
away from the black car bomb. Since the dangerous field of black car bomb is 
limited, the agents far away from it continue to move along their original 
paths. When the red car bomb occurs, these agents who are in the perilous 
field also begin to evacuate, while others just move in their original 
directions. Figure \ref{fig:13}(d) is the result at the end time (at the 
60$th$ frame). Animation details can be found in our supplemental video. 

\section{Discussion and Conclusion} 

Crowd behavior simulation under multi-hazard environment is a very 
challenging problem, and existing models with a single hazard cannot be 
applied to these cases directly. In this paper, we present a novel 
evacuation simulation method by modeling the generation and contagion of 
panic emotion under multi-hazard circumstances. First, we model multi-hazard 
environment by classifying hazards into different types based on their 
inherent characteristics and introducing the concept of perilous field for a 
hazard. Then, we propose a novel emotion contagion model to simulate the 
panic emotion evolving process in these situations. Finally, we introduce an 
emotional Reciprocal Velocity Obstacles(ERVO) model by augmenting the 
traditional RVO model with emotional contagion, which combines the panic 
emotion impact and local avoidance together for the first time. By comparing 
our simulation results with the ground-truth data and applying our algorithm 
in different virtual environments, 
our experiment results show that the overall 
approach is robust and can better generate realistic crowds as well as the panic 
emotion dynamics in a crowd in various multi-hazard environments.

There are still several limitations in our current work. The first one is 
that our current method relies on some important assumptions, such as
all agents in our scenario are treated equally in the face of hazards except the different 
personalities, thus they can perceive the danger level and be affected by the hazards once 
he/she enter into the influence radius of them. Besides that, safe exits 
chosen in the simulation environment in advance, especially in the office building 
situations, where the doors are chosen as the sole exit for each room. In real 
world, this is not very common. So we need to improve the sensing capability of the 
agents in an unknown multi-hazard scenario. The second is, in spite of considering 
agents personalities, expressiveness and susceptibility, diverse crowd movements are shown in our simulation 
results, many other complex personality traits and prior expertise may also affect the emotion changes and 
motion choices of each agent. In addition, the personality parameters in our emotional contagion model are 
set randomly to depict different agents, while it may not include all agents or some agents with special 
characters. Thus, more factors need to be considered. 
Furthermore, our method is sensitive to some key parameters, such as the strength of danger. 
In the future, we want to utilize a large number of surveillance video clips 
to calibrate and further improve our model.

 \ifCLASSOPTIONcompsoc
  % The Computer Society usually uses the plural form
   \section*{Acknowledgments}
    The authors would like to thank all the anonymous reviewers. This work was supported by National Natural Science Foundation of China under Grant Number 61672469, 61772474, 61822701, 61872324.
 \else
  % regular IEEE prefers the singular form
   \section*{Acknowledgment}
 \fi

% Can use something like this to put references on a page
% by themselves when using endfloat and the captionsoff option.
\ifCLASSOPTIONcaptionsoff
  \newpage
\fi

% trigger a \newpage just before the given reference
% number - used to balance the columns on the last page
% adjust value as needed - may need to be readjusted if
% the document is modified later
%\IEEEtriggeratref{8}
% The "triggered" command can be changed if desired:
%\IEEEtriggercmd{\enlargethispage{-5in}}

% references section

% can use a bibliography generated by BibTeX as a .bbl file
% BibTeX documentation can be easily obtained at:
% http://www.ctan.org/tex-archive/biblio/bibtex/contrib/doc/
% The IEEEtran BibTeX style support page is at:
% http://www.michaelshell.org/tex/ieeetran/bibtex/
%\begin{figure}
%  \centering
  % Requires \usepackage{graphicx}
%  \includegraphics[width=]{}\\
%  \caption{}\label{}
%\end{figure}
\bibliographystyle{IEEEtran}
% argument is your BibTeX string definitions and bibliography database(s)
\bibliography{bare_jrnl_compsoc}
\end{document}